\documentstyle[12pt]{article}

\def\hybrid{\topmargin 0pt      \oddsidemargin 0pt
        \headheight 0pt \headsep 0pt

       \textwidth 6.5in        
       \textheight 9in         
        \marginparwidth 0.0in
        \parskip 5pt plus 1pt   \jot = 1.5ex}
\catcode`\@=11
\def\marginnote#1{}

\newcount\hour
\newcount\minute
\newtoks\amorpm
\hour=\time\divide\hour by60
\minute=\time{\multiply\hour by60 \global\advance\minute by-\hour}
\edef\standardtime{{\ifnum\hour<12 \global\amorpm={am}%
        \else\global\amorpm={pm}\advance\hour by-12 \fi
        \ifnum\hour=0 \hour=12 \fi
        \number\hour:\ifnum\minute<10 0\fi\number\minute\the\amorpm}}
\edef\militarytime{\number\hour:\ifnum\minute<10 0\fi\number\minute}

\def\draftlabel#1{{\@bsphack\if@filesw {\let\thepage\relax
   \xdef\@gtempa{\write\@auxout{\string
      \newlabel{#1}{{\@currentlabel}{\thepage}}}}}\@gtempa
   \if@nobreak \ifvmode\nobreak\fi\fi\fi\@esphack}
        \gdef\@eqnlabel{#1}}
\def\@eqnlabel{}
\def\@vacuum{}

\def\draftmarginnote#1{\marginpar{\raggedright\scriptsize\tt#1}}

\def\draftlabel#1{{\@bsphack\if@filesw {\let\thepage\relax
   \xdef\@gtempa{\write\@auxout{\string
              \newlabel{#1}{{\@currentlabel}{\thepage}}}}}\@gtempa
   \if@nobreak \ifvmode\nobreak\fi\fi\fi\@esphack}
        \gdef\@eqnlabel{#1}}
\def\@eqnlabel{}
\def\@vacuum{}
\def\draftmarginnote#1{\marginpar{\raggedright\scriptsize\tt#1}}

\def\draft{\oddsidemargin -.5truein
        \def\@oddfoot{\sl preliminary draft \hfil
        \rm\thepage\hfil\sl\today\quad\militarytime}
        \let\@evenfoot\@oddfoot \overfullrule 3pt
        \let\label=\draftlabel
        \let\marginnote=\draftmarginnote
   \def\@eqnnum{(\theequation)\rlap{\kern\marginparsep\tt\@eqnlabel}%
\global\let\@eqnlabel\@vacuum}  }

\def\numberbysection{\@addtoreset{equation}{section}
        \def\theequation{\thesection.\arabic{equation}}}

\def\underline#1{\relax\ifmmode\@@underline#1\else
        $\@@underline{\hbox{#1}}$\relax\fi}

\def\titlepage{\@restonecolfalse\if@twocolumn\@restonecoltrue\onecolumn
     \else \newpage \fi \thispagestyle{empty}\c@page\z@
        \def\thefootnote{\fnsymbol{footnote}} }

\def\endtitlepage{\if@restonecol\twocolumn \else  \fi
        \def\thefootnote{\arabic{footnote}}
        \setcounter{footnote}{0}}  
\relax

\numberbysection
\hybrid

\def\beq{\begin{equation}}
\def\eeq{\end{equation}}
\def\p{\partial}
\def\G{\Gamma}
\def\g{\gamma}
\def\s{\sigma}

\def\bea{\begin{eqnarray}}
\def\eea{\end{eqnarray}}
\def\R{{\rm res}}
\newtheorem{th}{Theorem}

\newtheorem{lem}{Lemma}

\begin{document}

\begin{titlepage}

\title{The periodic and open Toda lattice}

\author{I.Krichever \thanks{Columbia University, 2990 Broadway,
New York, NY 10027, USA and Landau Institute
for Theoretical Physics, Kosygina str. 2, 117940 Moscow, Russia; e-mail:
krichev@math.columbia.edu. Research is supported in part by National Science
Foundation under the grant DMS-98-02577 and by CDRF Award RP1-2102}
\and
K.L. Vaninsky \thanks{Courant Institute New York  University
251 Mercer Street
NYC, NY 10012, vaninsky@cims.nyu.edu.
The work of K.V. is partially supported  by NSF grants DMS-9501002 and
DMS-9971834.} }
\date{October 19, 2000}
\maketitle

\begin{abstract}
We  develop algebro-geometrical approach for the open Toda lattice. 
For a finite Jacobi matrix  we introduce a singular reducible Riemann 
surface and associated Baker--Akhiezer functions. We provide new explicit 
solution of  inverse spectral problem for a finite Jacoby matrix.
For the Toda lattice equations  we obtain the   explicit form of the 
equations of motion, the symplectic structure and Darboux coordinates. 
We develop similar   approach  for 2D open Toda. Explaining some the 
machinery we also make contact with the periodic case.
\end{abstract}

\end{titlepage}
\newpage

\section {Introduction}
Until now the methods of integration for  periodic and open Toda lattice
were  absolutely unrelated to each other. The periodic Toda,  is a
Hamiltonian system of $N$--particles with the Hamiltonian
$$
H=\sum\limits_{k=1}^{N} {p_k^2\over 2} + \sum\limits_{k=1}^{N}e^{q_k-q_{k+1}},
\quad\quad\quad q_{N+1}=q_1+I_0,\quad\quad\quad I_0=const.
$$
The solution of  equations of motion  was obtained by Krichever  using  general
algebro--geometrical approach based on the concept of Baker--Akhiezer function
\cite{K1}. The solution was expressed in terms of theta functions
of the spectral curve, associated with the auxiliary spectral problem for
an infinite periodic Jacobi matrix. For the   open Toda  with the Hamiltonian
$$
H=\sum\limits_{k=1}^{N} {p_k^2\over 2}+\sum\limits_{k=1}^{N-1}e^{q_k-q_{k+1}}
$$
the solution of equations of motion
was obtained by Moser \cite{MO}. Moser reduced the original problem of
integration   to the inverse spectral problem for a finite Jacoby matrix.
The latter was solved  by Stieltjes \cite{S}  more than a hundred years ago
using continuous fractions.

Though the Hamiltonian of open Toda lattice  can be obtained from the
periodic one as the limit $I_0\to \infty$, the methods of solution are
different. This is  due to the fact that  auxiliary spectral problem
associated with a finite Jacoby matrix, unlike  the spectral problem
for the periodic infinite Jacoby matrix, does not determine  a natural 
spectral curve.
The first attempt to singularize smooth spectral curve of the periodic  Toda to
obtain  a spectral curve for  the open Toda chain  goes back to seventies 
\cite{mc}.
Recently, the  interest in this problem was revived \cite{mar,bm} 
due to connections of the  Toda lattice with Seiberg-Witten theory of 
supersymmetric $SU(N)$ gauge theory \cite{gor}. (A relatively complete list
of references on new insight upon the role of integrable systems in 
Seiberg-Witten theories \cite{SW1}-\cite{han} can be found in 
\cite{donagi}-\cite{dhoker} and books \cite{mar,bk}.)

The spectral curve proposed in \cite{mar,bm} for the open
Toda lattice is determined by the equation
\beq
w_0=P(E)=\prod\limits_{k=1}^{N} (E-E_k)
\eeq
and can be considered as the   limit when  $I_0\longrightarrow \infty$
of the hyper-elliptic  spectral curve
\beq\label{sp0}
w_0+{\Lambda^2\over w_0}=P(E),\quad\quad\quad \Lambda^2=e^{-I_0}
\eeq
for the periodic case.

From algebro--geometrical  viewpoint the limit
$\Lambda\longrightarrow 0$ of hyper-elliptic curve leads to a singular
curve, which is two copies of rational curve, corresponding to two sheets
of hyper-elliptic curve, glued together at  $N$ points $E_k$.
It is well known that Baker-Akhiezer functions introduced originally for
smooth algebraic curves, are also an important tool of integration in the 
limiting case of singular curves. Multi-soliton and rational solutions of 
integrable equations can be obtained within this approach (see in \cite{K3}).
The main goal of the present paper is to demonstrate that algebro--geometrical
approach based on the concept of Baker--Akhiezer function can be used in the
case of {\it reducible} singular algebraic curves. As an outcome,
we provide a solution to the inverse spectral problem for a finite Jacoby
matrix which is different from the  classical Stieltjes' solution.

In the recent papers by Krichever and Phong, \cite{KP1,KP2},
the new approach using Baker-Akhiezer functions for construction of
Hamiltonian theory of soliton type equations was developed. It provides
a universal scheme for construction of angle-action type variables
for these equations. In Section 4 we illustrate these ideas, and show that,
the case of the open Toda lattice with  singular curve can be treated 
similarly.

Finally, in section 5 we indicate how the ideas introduced in this paper can be applied to integration of open two-dimensional Toda lattice.

\section{Periodic Toda lattice.}

To begin with, let us first briefly recall algebro-geometric solution
of the periodic Toda lattice. That will help us later to clarify
algebro-geometric origin of our approach to the open Toda lattice.

Most of the material is standard and details can be found
in \cite{K2,T}. It is convenient to consider the periodic
Toda lattice system as a subsystem of an infinite lattice.
As it was found in \cite{F,M}, the equations of motion
\beq \label{eq}
\dot q_k= p_k, \ \ \dot p_k=
-e^{q_k- q_{k+1}}+e^{q_{k-1}- q_{k}}.
\eeq
$k=...,-1,0,1,...$ are equivalent to the Lax equation $\dot L=[A,L]$ for
auxiliary linear operators, where $L$ and $A$ are difference operators
\beq\label{laxL}
(L\psi)_n=c_n \psi_{n+1} +v_n \psi_n +c_{n-1} \psi_{n-1}
\eeq
\beq\label{laxA}
(A\psi)_n={c_n\over 2} \psi_{n+1} -{c_{n-1}\over 2} \psi_{n-1}
\eeq
with coefficients
\beq\label{cv}
c_k=e^{(q_k-q_{k+1})/2}, \quad \quad \quad v_k=-p_k.
\eeq
Let $q_n$ satisfy the constraint $q_n+I_0 =q_{n+N}$ which is invariant
with respect to (\ref{eq}). The corresponding $L$ and $A$ operators
are periodic ones: $c_n=c_{n+N},\ v_n=v_{n+N}$. The Floquet-Bloch solution
is the solution of the periodic Schr\"odinger equation
\beq\label{bl}
(L\psi)_n=c_n \psi_{n+1} +v_n \psi_n +c_{n-1} \psi_{n-1}=E\psi_n
\eeq
such that
\beq\label{w}
\psi_{n+N}=w\psi_n.
\eeq
Let $T(E)$ be a restriction of the monodromy operator
$(Tf)_n=f_{n+N}$ on the invariant space of solutions to the equation
$(L\psi)_n=E\psi_n$.
The solutions $\varphi$ and $\theta$ of the Schr\"odinger equation,
normalized  such that $\varphi_0=1,\ \varphi_1=0$ and $\theta_0=0,\ \theta_1=0$
define a basis in which the monodromy operator have the form:
\beq
T(E)=\left[\begin{array}{cc}  \varphi_N&\theta_N\\ \varphi_{N+1}&\theta_{N+1}
\end{array}\right].
\eeq
Then pairs of complex numbers $(w,E)$ for which there
is a common solution of equations (\ref{bl},\ref{w}) are defined by
characteristic equation
\beq\label{spec}
R(w,E)= \det (w-T(E))=w^2-2\Delta(E)w+1=0.
\eeq
The hyperelliptic Riemann surface $\Gamma$  defined by equation (\ref{spec}) is called
a {\it spectral curve} of the periodic Schr\"odinger operator.
A point on the curve we denote by $Q=(w,E)$.

To make contact with the open Toda case it is useful to introduce a different
representation of $\Gamma$.  Consider the operator $L(w)$ which is defined
as a restriction of the infinite-dimensional operator $L$ on $N$-dimensional
space of functions that satisfy equation (\ref{w}). The corresponding
matrix has the form
\beq\label{lw}
L(w)=\left[\begin{array}{ccccc}
v_0      & c_0       & 0         &\cdots        & w^{-1}c_{N-1}\\
c_0      & v_1       & c_1       &\cdots        & 0\\
0        &  \cdot    & \cdot     &\cdot         & 0\\
0    &  \cdots         & c_{N-3}         & v_{N-2}     & c_{N-2}\\
wc_{N-1}       & \cdots    & 0         & c_{N-2}      & v_{N-1}
\end{array}\right]
\eeq
From (\ref{lw}) it follows that
\beq\label{spec1}
\det (E-L(w))=
P(E)-\Lambda w-\Lambda w^{-1}=0,\ \
\Lambda=(-1)^N\prod_{n=1}^N c_n=(-1)^Ne^{-I_0/2}.
\eeq
Coefficients of the monic polynomial $P(E)$
\beq\label{u}
P(E)=E^N+\sum_{i=0}^{N-1} u_iE^i
\eeq
are polynomial functions of the coefficients $c_n,v_n$ of $L$.
For example,
\beq\label{ui} u_{N-1}=-\sum_{n=1}^{N} v_n,\ \
u_{N-2}=\left(\sum_{0<i<j}^Nv_iv_j-\sum_{i=1}^Nc_i^2\right).
\eeq
Note, that $P=2\Lambda\Delta(E)$. Change of the variable $w_0=
\Lambda w$ transforms (\ref{spec1}) into the desired formula (\ref{sp0}).

The equation $\partial_w R(w,E)=0$  for  branch points  of $\Gamma$
leads to $\Delta^2(E)=1$. It  has
$2N$ real roots which are points of periodic/anti-periodic spectrum for $L$,
according to  the sign of $\Delta(E)=\pm1$ at these points. For a generic
configuration of particles these roots are distinct
$E_{1}<E_{2}<\cdots<E_{2N}$ and the curve $\G$ is smooth and has  genus $N-1$.

To make a model of the curve $\Gamma$ we take two copies of the complex plane
and make $N$ cuts along the bands $[E_{1},E_{2}],\ldots,[E_{2N-1},E_{2N}]$.
Then glue two cut-planes together.

The multiplier is holomorphic in the affine part of the curve with
pole/zero of degree $N$ at the infinity $P_+/P_-$. The multivalued function
of quasimomentum $p(Q)$ is introduced by the formula $w=e^{Np}$. It has  the
asymptotic expansion
$$
\pm p(E)= \log E-\sum\limits_{k=0}^{\infty} {H_k\over E^k},\quad\quad\quad\quad
E=E(Q),\;\;\; Q\rightarrow  (P_{\pm}).
$$
The coefficients are standard integrals of the periodic Toda
$$
H_0 =-{I_0\over 2N},\quad\quad\quad
H_1 ={1\over N} \sum\limits_{k=1}^{N} v_k, \quad\quad\quad
H_2 ={1\over N} \sum\limits_{k=1}^{N} {v_k^2\over 2} +
\sum\limits_{k=1}^{N} c_k^2.
$$
The explicit
formula for the Floquet-Bloch solution
\beq\label{bl1}
\psi_n(Q)=\varphi_n(E) + {w(Q)- \varphi_N(E)\over \theta_N(E)} \theta_n(E).
\eeq
implies that
\begin{lem} (i)
The Floquet-Bloch solution, normalized by the condition
$\psi_0=1$
becomes single-valued on the Riemann surface $\Gamma$. It has a single pole
$\gamma_k$ on each real oval $a_k$, $k=1,\ldots,N-1$, which is a preimage
of the $[E_{2k}, E_{2k+1}]$.

(ii) In the vicinity of infinities  the Floquet solution  has the asymptotics
$$
\psi_n(E)=E^{\pm n}e^{\pm (q_n-q_0)/2}\left(1+\sum_{s=1}^{\infty}
\chi_s^{\pm}(n)E^{-s}\right),
$$
where $E=E(Q), \;\quad Q\to (P_{\pm})$.
\end{lem}
Equation (\ref{bl1}) implies that projections $E(\g_s)$ of the poles of $\psi$
are zeros of the polynomial $\theta_N(E):\, \theta_N(E(\g_s))=0$. Therefore,
$\theta_n(E(\g_s))$ satisfies zero boundary conditions, and $E(\g_s)$ are
points of the Dirichlet spectrum for the Schr\"odinger operator.

The map which associates for $N$-periodic operator $L$ with
coefficients $\{c_n,v_n\}$ a spectral curve $\G$, and
the divisor $D=\{\g_1,\ldots,\g_{N-1}\}$ of poles of the Floquet-Bloch solution.
\beq
\{c_n,v_n\}\to \{\Gamma,D\},\label{map}
\eeq
is referred as {\it direct spectral transform}.

The spectral curve is defined by the variable $\Lambda$ , and
$N$ coefficients $u_i$ of the polynomial
$P(E)$. Therefore, a space of the spectral data has the same
dimension $2N$ as the space of $N$-periodic operators.
It is a fundamental fact in the theory
that the map (\ref{map}) is a bijective correspondence of generic points.
We shall refer to the reverse construction
\beq \label{map1}
\{\Gamma,D\}\longmapsto\{c_n,v_n\},
\eeq
which recaptures the dynamical variables $\{c_n,v_n\}$
from the geometric data $\{\Gamma,D\}$,
as {\it the inverse problem}.
As usual in the algebro-geometric theory of solitons,
it will be based on the construction of a
Baker-Akhiezer function.

\begin{lem}\cite{K1} Let $\G$ be a smooth hyperelliptic curve defined
by equation (\ref{spec1}). Then for a generic set of $(N-1)$ points
$\g_s\in \G$ there exists a unique (up to a sign) function $\Psi_n(t,Q)$ such
that:

(i) it is meromorphic function on $\G$ outside infinities $P_{\pm}$
and has at most simple poles at the points $\g_s$.

(ii) In the vicinity of infinities  the Floquet solution  has the asymptotics
$$
\Psi_n(E)=E^{\pm n}e^{\pm Et/2}b_n^{\pm 1}(t)\left(1+\sum_{s=1}^{\infty}
\chi_s^{\pm}(n,t)E^{-s}\right),
$$
where $E=E(Q), \;\quad Q\to (P_{\pm})$.
\end{lem}
Uniqueness of $\Psi$ immediately implies the following theorem.
\begin{th}
The Baker-Akhiezer function $\Psi_n(t,Q)$ associated with the spectral data
$\{\G,D\}$ satisfies equations
\beq\label{LA}
(L\Psi)_n(t,Q)=E\Psi_n(t,Q), \ \ (\p_t-A)\Psi_n(t,Q)=0.
\eeq
where the operators $L$ and $A$ have the form (\ref{laxL})
and (\ref{laxA}) with coefficients defined by the formulae
$$
c_n(t)=b_n(t)b_{n+1}^{-1}(t), \ \ v_n=\chi_1^{+}(n,t)-\chi_1^{+}(n+1,t).
$$
These coefficients are $N$-periodic functions of the variable $n$.
\end{th}
From the Lax equation, which is a compatibility condition for
({\ref{LA}) we obtain that
\beq\label{tq}
\tilde q_n(t)=\ln b_n(t)
\eeq
is a solution of the periodic Toda lattice normalized by the
condition $\tilde q_0(0)=0$.
Note, that for $t=0$ the Baker-Akhiezer function has the same
analytical properties as the Floquet-Bloch solutions. Therefore,
$\Psi_n(0,Q)=\psi_n(Q)$, and using the spectral data $\{\G,D\}$
associated with $\{c_n,v_n\}$ we solve the Cauchy problem for the periodic
Toda lattice.

This solution can be written explicitly in terms of
the Riemann $\theta$-function associated with matrix of $b$-periods of
a normalized holomorphic differentials  \cite{K2}.
\beq\label{for}
q_n(t)=q_0(0)+{n\over N} I_0+\ln\left(\theta((n-1)U+Vt+Z)\theta(Z)\over
\theta(nU+Vt+Z)\theta(Z-U)\right)
\eeq
Here $U$ and $V$ are $(N-1)$-dimensional vectors that are periods
of certain meromorphic differentials on $\G$, and $Z$ is Abel transform
of the divisor $D$.

We finish this brief account of solution of the periodic Toda lattice
by the following remark. The Lax equation implies that if the dynamical
variables $\{c_n(t),v_n(t)\}$ evolve according to the Toda lattice
equations then the spectral curve of the Schr\"odinger operator is
{\it time-independent}.  At the same time the poles, of the Floquet-Bloch
solution normalized by the condition $\psi_0=1$ do depend on $t$. From
(\ref{LA}) it follows that $\psi_n(t,Q)$ and the Baker-Akhiezer function
are proportional to each other. Therefore,
$$\psi_n(t,Q)=\Psi_n(t,Q)\Psi_0^{-1}(t,Q)$$
Hence, the divisor $D(t)$ can be identified with the divisor of zeros of
the function $\Psi_0(t,Q)$. An image of this divisor under the Abel transform
evolves linearly, $Z(t)=Z+Vt$.

\section{Open Toda lattice}

Equations of motion for the open N-particle Toda lattice
\begin{eqnarray}
\dot q_k& =&  p_k, \quad\quad \quad\quad\quad \quad\quad  \quad\quad\quad \quad\quad \quad\quad  \quad\quad\quad k=0,...,N-1, \nonumber\\
\dot p_k & = & -e^{q_k- q_{k+1}}+e^{q_{k-1}- q_{k}} \quad \quad \quad\quad\quad \quad\quad
\quad\quad\quad k=1,...,N-2 ,\nonumber\\
\dot p_0  &= &-e^{q_0- q_{1}},  \quad\quad\quad \quad \dot p_{N-1}  = e^{q_{N-2}- q_{N-1}}, \nonumber
\end{eqnarray}
have Lax representation
$\p_tL_0=[A_0,L_0]$ where $L_0$ and $A_0$ are finite-dimensional Jacobi matrices
\beq\label{LAop}
L_0=\left[\begin{array}{ccccc}
v_0      & c_0       & 0         &\cdots        & 0\\
c_0      & v_1       & c_1       &\cdots        & 0\\
\cdot        &  \cdot    & \cdot     &\cdot         & \cdot\\
0    &  \cdots         & c_{N-3}         & v_{N-2}     & c_{N-2}\\
0       & \cdots    & 0         & c_{N-2}      & v_{N-1}
\end{array}\right],\ \
A_0={1\over 2}\left[\begin{array}{ccccc}
0      & c_0       & 0         &\cdots        & 0\\
-c_0      &   0     & c_1       &\cdots        & 0\\
\cdot        &  \cdot    & \cdot     &\cdot         & \cdot\\
0    &  \cdots         & -c_{N-3}         & 0     & c_{N-2}\\
0       & \cdots    & 0         & -c_{N-2}      & 0
\end{array}\right]
\eeq
where the variables $c_n,\, v_n$ are defined by formulae (\ref{cv}),
as in the periodic case.

Equations of motion are invariant with respect to the shift
$v_n\to v_n+const$. Therefore, without loss of generality we will
assume from now on that $v_n$ are normalized by the constraint
\beq\label{i1}
\sum_{n=0}^{N-1}v_n=0
\eeq

As it was mentioned in the introduction, the spectral problem for $L$
does not contain spectral parameter. Let us introduced another
auxiliary spectral problem, which does contain spectral parameter.
The very same equations of motion are equivalent
to the Lax equation for the following finite-dimensional operators:
\beq \label{law}
L(w)=\left[\begin{array}{ccccc}
v_0      & c_0       & 0         &\cdots        & 0\\
c_0      & v_1       & c_1       &\cdots        & 0\\
\cdot        &  \cdot    & \cdot     &\cdot         & \cdot\\
0    &  \cdots         & c_{N-3}         & v_{N-2}     & c_{N-2}\\
wc_{N-1}       & \cdots    & 0         & c_{N-2}      & v_{N-1}
\end{array}\right],
\eeq
\beq\label{law1}
A(w)={1\over 2}\left[\begin{array}{ccccc}
0      & c_0       & 0         &\cdots        & 0\\
-c_0      &   0     & c_1       &\cdots        & 0\\
\cdot        &  \cdot    & \cdot     &\cdot         & \cdot\\
0    &  \cdots         & -c_{N-3}         & 0     & c_{N-2}\\
wc_{N-1}       & \cdots    & 0         & -c_{N-2}      & 0
\end{array}\right]\ ,
\eeq
where
\beq\label{c}
c_{N-1}=e^{(q_{N-1}-q_0)/2}=\prod_{n=0}^{N-2}c_n^{-1}.
\eeq
Coordinates $\psi_n(E),\ n=0,\ldots, N-2$, of an eigenvector $\psi$,\
$
L(w)\psi(E)=E\psi(E)
$
can be found reccurently from equations
\begin{eqnarray}\label{21}
E\psi_0&=&c_0\psi_1+v_0\psi_0,\nonumber \\
E\psi_n&=&c_n\psi_{n+1}+v_n\psi_n+c_{n-1}\psi_{n-1}, \ \ n=1,\ldots,N-2.
\end{eqnarray}
If we take $\psi_0=1$, then (\ref{21}) implies that $\psi_n(E)$ is a polynomial
of degree $n$
\beq\label{ps0}
\psi_n(E)=\sum_{i=0}^n b_i(n)E^j,\quad\quad \ b_n(n)=e^{(q_n-q_0)/2}.
\eeq
The last equation
\beq
E\psi_{N-1}=c_{N-1}w\psi_{0}+v_{N-1}\psi_{N-1}+c_{N-2}\psi_{N-2}
\eeq
defines $w=w(E)$. It is a polynomial of degree $N$. From (\ref{c}) it follows
that $w(E)$ is a monic polynomial. It can be also found from the characteristic
equation
\beq\label{P}
R_0(w,E)=\det (E-L(w))=P(E)-w=0, \ \ P(E)=E^N+\sum_{i=0}^{N-2}u_iE^i.
\eeq
which  determines the components of the spectral curve. For $w=0$ the matrix
$L(w)$ coincides with $L_0$.  Therefore, zeros $E_k$ of the polynomial $w(E)$
\beq\label{www}
w(E)=\prod_{k=1}^N(E-E_k)
\eeq
are eigenvalues of the matrix $L_0$. The  spectral  curve $\Gamma_0$ is
singular algebraic curve which is  obtained by gluing  to each other two copies
of the $E$--plane along the set of N points  $E_k$.

Let us introduce now a solution $\varphi=\{\varphi_n(E)\}$ of the adjoint equation
$L^T(w)\varphi=E\varphi$. If we normalize $\varphi$ by the condition $\varphi_{N-1}=1$,
then the equations
\begin{eqnarray}
E\varphi_{N-1}&=&c_{N-2}\varphi_{N-2}+v_{N-1}\varphi_{N-1}, \nonumber \\
E\varphi_n\ &=&c_n\varphi_{n+1}+v_n\varphi_n+c_{n-1}\varphi_{n-1}, \ \ n=N-2,\ldots,1.
\nonumber
\end{eqnarray}
reccurently define $\varphi_n(E), \;\;n=0,\ldots, N-2$. It is a polynomial of
degree $N-1-n$ with the leading coefficient
$$
\varphi_n(E)=e^{(q_{N-1}-q_n)/2}E^{N-1-n}+O(E^{N-2-n}).
$$
The  last equation
$$
E\varphi_0\ =c_0\varphi_{1}+v_0\varphi_0+ wc_{N-1}\varphi_{N-1},
$$
which determines $w$  leads to  the same formula (\ref{www}).

Let $\g_s$ be roots of the equation $\varphi_0(E)=0$.
As we will show,   they interlace $N$ roots
of the determinant of $L_0$, i.e.
\beq\label{30}
E_1<\g_1<E_2<\ldots<E_{N-1}<\g_{N-1}<E_N.
\eeq
Coordinates
of an eigenvector $\psi^{\s}(E)=\psi_n^{\s}(E)$ for the matrix $L^T(w)$,
normalized by the condition $\psi_0^{\s}=1$, are equal to $\psi^{\s}_n=
\varphi_0^{-1}(E)\varphi_n(E)$, and therefore, have the form
\beq\label{ps}
\psi_n^{\s}={\sum_{j=0}^{N-1-n}b_j^{\s}(n)E^{j}\over
\prod_{s=1}^{N-1}(E-\g_s)},\quad\quad \ b_{N-1-n}^{\s}(n)=e^{(q_0-q_n)/2}.
\eeq
At $w=0$ matrices $L(w)$ and $L^T(w)$ coincides. Therefore, their
eigenvectors are proportional to each other. Due to our choice of
normalization, we conclude that
\beq\label{gl}
\psi_n(E_k)=\psi_n^{\s}(E_k).
\eeq
Due to (\ref{gl}), a pair of functions $\psi_n(E),\psi_n^{\s}(E)$
can be considered as a single-valued function on  $\G_0$.
This is the Baker-Akhiezer function for  the spectral curve $\G_0$.

To prove (\ref{30}) we represent the  solution $\varphi$ using truncated matrices
\beq\label{LK}
L_k=\left[\begin{array}{ccccc}
v_k      & c_k       & 0         &\cdots        & 0\\
c_k      & v_{k+1}       & c_{k+1}   &\cdots        & 0\\
\cdot        &  \cdot    & \cdot     &\cdot         & \cdot\\
0    &  \cdots         & c_{N-3}         & v_{N-2}     & c_{N-2}\\
0       & \cdots    & 0         & c_{N-2}      & v_{N-1}
\end{array}\right],
\eeq
We define $\det(E-L_{N+1})=0,\quad \det(E-L_{N})=1$. Using the reccurence relations
$$
\det(E-L_k)=(E-v_k)\det(E-L_{k+1})-c_{k}^2 \det(E-L_{k+2}),
$$
it is easy to check, that
$$
\varphi_n(E)={\det(E-L_{n+1}) \over c_n .... c_{N-2}}.
$$
From this we see that $\gamma_s$ are the spectrum of the matrix $L_1$ which is obtained from $L_0$
by canceling the first raw and the first column. Now classical theorem of Sturm \cite{GK} implies
(\ref{30}). By analogy with the periodic case we call the points $\gamma_s$ the divisor $D_0$.
The points of $D_0$ move between $E_k$ under the Toda flow, but never change a sheet, contrary to the
periodic case.

As in the periodic case, a map which associates to dynamical variables
a set of spectral data $\Gamma_0,\, D_0$ of the same dimension is a bijective correspondence
\beq\label{map3}
\{c_0,\ldots,c_{N-2},v_0,\ldots,v_{N-1}\}\longleftrightarrow \{\G_0,D_0\}.
\eeq
In fact, the gluing conditions (\ref{gl}), and the normalization
\beq\label{nor1}
b_{n}(n)b^{\s}_{N-1-n}(n)=1,
\eeq
which follows from (\ref{ps0}) and (\ref{ps}), allow to reconstruct $\psi_n$ and
$\psi_n^{\s}$.

For each $n$ the gluing conditions are equivalent to $N$ linear
equations for $N+1$ unknown coefficients $b_i,\,  b_j^{\s}$.
If we define the vector with $N+1$ components
$B_j=b_j,\;j=0,\ldots,n;\; B_j=b_{j-n-1}^{\s},\; j=n+1,\ldots,N$ then the matrix
of the corresponding linear system has the form
\begin{eqnarray}\label{M}
M_{ij}(n)&=&r_i E_i^j,\quad\quad \quad\quad \quad \quad j=0,\ldots, n \nonumber\\
M_{ij}(n)&=& -E_i^{j-n-1}, \quad\quad \quad\quad j=n+1,\ldots,N,
\end{eqnarray}
where $r_i=\prod_{s=1}^{N-1} (E_i-\gamma_s),\; i=1,\ldots,N$. Let
$\Delta_j(n)$
be the $j$--th minor obtained by  canceling  the $j$--th column.
The coefficients $B$'s which satisfy the normalization condition are
\beq\label{f20}
B_j(n)={\Delta_j(n)\over \sqrt{\Delta_{n}(n)\, \Delta_N(n)}},
\quad\quad\quad\quad\quad j=0,\ldots,N.
\eeq
Since $B_{n}(n)=e^{(q_n-q_0)/2}$, then
\beq\label{f11}
q_n-q_0=2\ln B_{n}(n)=\ln {\Delta_{n}(n)\over \Delta_{N}(n)},\ \ n=0,\ldots,N-1.
\eeq
Of course, there are sets of complex numbers $\{E_k,\g_s\}$,
such that the right hand side of
(\ref{f11}) is singular.
\begin{lem} Let  the spectral data $\{E_k,\g_s\}$ be real
and satisfy conditions (\ref{30}), then
\beq\label{sg}
\Delta_n(n)\Delta_{N}(n)\neq 0.
\eeq
\end{lem}
Indeed, let $\tilde \psi_n$ and $\tilde \psi_n^{\s}$ be
the unnormalized solution of the gluing conditions with the coefficients
$b_i,\ b_j^{\s}$, corresponding to $\tilde B_j(n)=\Delta_j(n)$.
The product $\tilde \psi_n\tilde \psi_n^{\s}$ is a rational function of  the form
$$\tilde \psi_n\tilde \psi_n^{\s}={F(E)\over \prod_{s=1}^{N-1}(E-\g_s)},
\ \ F=\Delta_n(n)\Delta_{N}(n)E^{N-1}+O(E^{N-2}).
$$
If the spectral data are real, then $\tilde \psi_n(E)$ and $\tilde \psi_n^{\s}(E)$
are real for real values of $E$. Due to the gluing conditions
the product of these functions at $E=E_k$ is positive. Therefore, the product
$\tilde \psi_n\tilde \psi_n^{\s}$ has
at least one zero on each interval $[E_k,E_{k+1}]$, because it has a pole
there. Hence, degree of the polynomial $F$ can not be less then $N-1$.
That implies (\ref{sg}).

\begin{th} Let $\psi_n(E), \psi_n^{\s}(E)$ be functions
associated with spectral data $E_k,\g_s$. Then they satisfy equations
$L(w)\psi=E\psi$ and $L^T(w)\psi^{\s}=E\psi^{\s}$,
where
\beq\label{c11}
c_n^2={\Delta_{n}(n)\Delta_N(n+1)\over \Delta_{n+1}(n+1)\Delta_{N}(n)},
\ \ n=0,\ldots, N-2,
\eeq
and
\begin{eqnarray}
v_n&=&{\Delta_{n-1}(n)\over \Delta_n(n)}-
{\Delta_{n}(n+1)\over \Delta_{n+1}(n+1)},\  \ n=1,\ldots N-2;\nonumber\\
v_0&=&-{\Delta_{0}(1)\over \Delta_{1}(1)},\ \ \ \
v_{N-1}={\Delta_{N-2}(N-1)\over \Delta_{N-1}(N-1)}.\nonumber
\end{eqnarray}
\end{th}
{\it Proof.} Functions $\psi_n, \psi_n^{\s}$ at $E\to \infty$ have the form
\begin{eqnarray}
\psi_n(E)&=&E^{n}e^{(q_n-q_0)/2}\left(1+\sum_{s=1}^{\infty}
\chi_s(n)E^{-s}\right),\label{ex1}\\
\psi_n^{\s}(E)&=&E^{-n}e^{(q_0-q_n)/2}\left(1+\sum_{s=1}^{\infty}
\chi_s^{\s}(n)E^{-s}\right)\ .\label{ex2}
\end{eqnarray}
Note, that
\beq\label{f21}
\chi_1(n)={b_{n-1}(n)\over b_n(n)},\ \
\chi_1^{\s}(n)={b^{\s}_{N-2-n}(n)\over b^{\s}_{N-1-n}(n)}+\sum_{s=1}^{N-1}\g_s.
\eeq
Here for $n=0$ and $n=N-1$ we formally put $b_{-1}(0)=b_{-1}^{\s}(N-1)=0$.

Let $\phi_n(E)$ and $\phi_n^{\s}(E)$ be coordinates of the vectors
\beq
\phi(E)=(L(w)-E)\psi(E), \ \ \phi^{\s}(E)=(L^T(w)-E)\psi^{\s}(E).
\eeq
If $c_n$ are defined by (\ref{cv}), and $v_n$ are defined by the equations
\beq
v_n=\chi_1(n)-\chi_1(n+1),\ n=0,\ldots,N-2;\
v_{N-1}=\chi_1(N-1),
\eeq
then (\ref{ex1}) implies that  $\phi$ near infinity has the form
$\phi_n=O(E^{n-1})$. Therefore, we conclude that $\phi_n$ is a polynomial
of degree at most $(n-1)$. The leading term of expansion for $\phi_n^{\s}$
near infinity is $O(E^{-n})$. Hence, $\phi_n^{\s}$ has the same form as
$\psi_n^{\s}$.  Functions $\phi_n$ and $\phi_n^{\s}$ satisfy the same gluing
conditions (\ref{gl}). They are equivalent to $N$ linear equations for
$N$ unknown
coefficients for corresponding polynomials. The matrix of this system
is obtained from the matrix $M$ (\ref{M}) by canceling the $n$-th column.
If this matrix is non degenerate, $\Delta_n(n)\neq 0$,
we conclude that $\phi=\phi^{\s}=0$.
Formulae (\ref{f20},\ref{f21}) allow to complete   proof of the  theorem.

Note, that the equation $(L^T-E)\psi^{\s}=0$ implies that
\beq
v_0=-\chi_1^{\s}(N-1),\ \ v_n=\chi_1^{\s}(n)-\chi_1^{\s}(n-1),\ n=1,\ldots,N-1,
\eeq
Therefore, we obtain the identities
\beq\label{100}
\chi_1(n+1)+\chi_1^{\s}(n)=-v_0,
\eeq
which will be used in the next section.

Solution of the Cauchy problem for the open Toda lattice can be obtained in
a way which is almost identical to the periodic case. First of all, we
defined time-dependent Baker-Akhiezer functions
$\psi_n(t,E),\ \psi_n^{\s}(t,E)$, as functions of the form
\begin{eqnarray}
\psi_n(t,E)&=&e^{Et/2}\left(\sum_{i=0}^n b_i(t,n)E^j\right),\\
\psi_n^{\s}(t,E)&=&e^{-Et/2}\left({\sum_{j=0}^{N-1-n}b_j^{\s}(t,n)E^{j}\over
\prod_{s=1}^{N-1}(E-\g_s)}\right),
\end{eqnarray}
that satisfy gluing (\ref{gl}) and normalization conditions (\ref{nor1}).

After that, without any change of arguments we obtain that they satisfy the
equations $(L(t,w)-E)\psi(t,E)=0,\ (L^T(t,w)-E)\psi^{\s}(t,E)=0$. In
the similar
way we get that these functions satisfy equations
$(\p_t-A(t,w))\psi(t,E)=0,\ (\p_t-A^T(t,w))\psi^{\s}(t,E)=0$.
As a result of compatibility condition we get that functions
$q_n(t)=q_0(0)+\ln b_n(t,n)$ are solutions of the open Toda lattice.

The matrix of the linear system for $b_i(t,n)$ and $b_j^{\s}(t,n)$ has the same
form as in (\ref{M}) with the only difference that  constants $r_i$ should be
replaced by $r_i(t)=r_ie^{E_kt}$. Therefore, we come to the following theorem.
\begin{th}
Let $\Delta_j(t,n)$ be the $j$-th minor obtained by canceling the $j$-th column
of the $N\times (N+1)$ matrix
\begin{eqnarray}
M_{ij}(t,n)&=&r_i E_i^je^{E_it},\quad\quad \quad\quad \quad j=0,\ldots, n
\nonumber\\
M_{ij}(t,n)&=& -E_i^{j-n-1}, \quad\quad \quad\quad j=n+1,\ldots,N,
\nonumber
\end{eqnarray}
where $r_i=\prod_{s=1}^{N-1}(E_i-\g_s)$.
Then the formula
$$
q_n(t)=q_0(0)+\ln{\Delta_{n}(t,n)\over \Delta_N(t,n)}
$$
gives a solution of the Cauchy problem for the open Toda lattice equations.
\end{th}

\section{Action-angle type variables}

The main goal of this section is to show that  Lax representation which 
depends  on the spectral parameter allows to apply for  Hamiltonian theory of 
open Toda
lattice an algebro-geometric approach proposed in \cite{KP1,KP2}. 
The main idea of
this approach is to introduce in a universal way two-form on a space of
auxiliary linear operators, written in terms of the operator itself and its
eigenfunctions. This form defines on proper subspaces a symplectic structure
with respect to which  Lax equation is Hamiltonian. Moreover, the
way to define the symplectic structure leads directly to construction of
Darboux coordinates.

In the case of consideration, we define such a form by the formula
\beq
\omega=\R_{\infty}\left[<\psi^+\delta L \wedge \delta \psi>+
<\delta\psi^+\wedge\delta L \psi>\right]{d\ln w\over <\psi^+\psi>}.
\label{omega}
\eeq
Here $\delta L$ is an operator-valued one-form on a space of coefficients
of the operator, i.e. in our case it is just a matrix with entries,
which are one-forms
\beq\label{dL}
\delta L(w)=\left[\begin{array}{ccccc}
\delta v_0      & \delta c_0       & 0         &\cdots        & 0\\
\delta c_0      &\delta v_1       &\delta c_1       &\cdots        & 0\\
\cdot        &  \cdot    & \cdot     &\cdot         & \cdot\\
0    &  \cdots         &\delta c_{N-3}         &\delta v_{N-2}
&\delta c_{N-2}\\
w\delta c_{N-1}       & \cdots    & 0         &\delta c_{N-2}
 & \delta v_{N-1}
\end{array}\right].
\eeq
An eigen-vector $\psi$ of $L$ can be seen as a vector-valued function on the
space of the coefficients of $L$, and therefore, its external
differential $\delta \psi$ is a vector-valued one-form. The co(row)-vector
$\psi^+$ is an eigen-vector of $L$ in the dual space, $\psi^+L=E\psi^+$, i.e.
$\psi^+=(\psi^{\s})^T$. At last, $ <f^+g>$ stands for pairing
of vector and co-vector
$$<f^+g>=\sum_{n=0}^{N-1}f_n^+g_n.$$

Now, it is necessary to add an important remark, which clarifies our
understanding of $\delta \psi$.
In (\ref{dL}) variation of $L$ was defined such that the spectral parameter
$w$ is fixed. An eigenvalue of $L$ can be seen locally as a function
$E=E(w)$ defined implicitly by (\ref{P}).
Therefore, though $\psi(E)$ is a polynomial in $E$, locally,
we consider it as a
function of $w$, $\psi(w)=\psi(E(w))$, and partial derivatives in
$$\delta =
\sum\left(\delta c_n{\p\over\p c_n}+\delta v_n{\p\over\p v_n}\right)$$
are taken for fixed values of $w$. A connection between $\delta$ and
variation $\delta_0$, which is defined if one keeps $E$ fixed, is just
a chain rule. Let $f(E)=f(E,c_n,v_n)$ be a function of the variable $E$
depending on $\{c_n,v_n\}$ as on external parameters. Then
\beq\label{de0}
\delta f(w)=\delta f(E(w))=\delta_0 f(E)+{df\over dE}\delta E(w).
\eeq
In particular, if we take $f=w$, then from (\ref{de0}) it follows that
$$0=\delta_0 w(E)+{dw\over dE}\delta E(w).$$
Therefore, (\ref{de0}) may be rewritten in the form
\beq\label{de}
\delta f(w)=\delta_0 f(E)-{df\over dw}\delta_0 w(E).
\eeq
This formula shows that variation $\delta$ leads to an appearance of poles
at zeros of $dw$, which are critical points of the polynomial $w(E)$,
where we, even locally, can not invert $w$ and introduce $E=E(w)$.

Our immediate goal is to find an explicit formula for $\omega$ in term of
dynamical variables.
\begin{lem}
Let $\omega$ be given (\ref{omega}). Then on symplectic leaves
of the space of dynamical variables $\{c_n, v_n\}$, defined by
the constraint $\sum_{n}v_n=const$, the form $\omega$ is equal to
\beq\label{qv}
\omega=2\sum_{n=0}^{N-1}\delta q_n\wedge \delta v_n.
\eeq
\end{lem}
{\it Proof.} In order to simplify slightly formulae, we consider
only the leaf $\sum_{n}v_n=0$. For all the other leaves a proof is going
in the same way.

At infinity the functions $\psi_n$ and $\psi^{\s}$ can be expanded
as Laurent series in the variable
$$k=w^{1/N}(E)=E+O(E^{-1}).$$
From (\ref{ex1},\ref{ex2}) we get
\begin{eqnarray}
\psi_n(k)&=&k^{n}e^{(q_n-q_0)/2}\left(1+\sum_{s=1}^{\infty}
\xi_s(n)k^{-s}\right)\label{ex3},\\
\psi_n^{\s}(k)&=&k^{-n}e^{(q_0-q_n)/2}\left(1+\sum_{s=1}^{\infty}
\xi_s^{\s}(n)k^{-s}\right) \label{ex4}.
\end{eqnarray}
Note, that in this expansion $\xi_1=\chi_1,\ \xi_1^{\s}=\chi_1^{\s}$.
Therefore, as before,
\beq\label{110}
v_n=\xi_1(n)-\xi_1(n+1), \ \xi_1(n+1)+\xi_1^{\s}(n)=-v_0\quad\quad\quad
n=0,\ldots,N-2,
\eeq
\beq\label{120}
\xi_1(N-1)=v_{N-1}.
\eeq
The variable $k$ is a constant with respect to our definition of $\delta$, and
therefore, an expansion of $\delta \psi$ can be obtained by taking variation
of coefficients in (\ref{ex3},\ref{ex4}).

Definition of $\delta L$ implies
$$
<\psi_n^+\delta L\wedge \delta \psi>=
\sum_{n=0}^{N-1}\left(\delta v_n\wedge \psi_n^+\delta \psi_n\right)+
\sum_{n=0}^{N-2}\left(\delta c_n\wedge \left(\psi^+_n\delta \psi_{n+1}+
\psi^+_{n+1}\delta \psi_{n}\right)\right)
$$
From (\ref{110},\ref{120}) it follows that at infinity
\beq
{d\ln w\over <\psi^+\psi>}=\left(1+v_0k^{-1}+O(k^{-2})\right){dk\over k}\ .
\nonumber
\eeq
Using expansions (\ref{ex3},\ref{ex4}) and equation (\ref{110}) we obtain
\beq
\omega=
-2\left[\sum_{n=0}^{N-1}\delta v_n\wedge \delta(q_n-q_0)+
\sum_{n=0}^{N-2}\delta(q_n-q_{n+1})\wedge \delta \xi_1(n+1)\right].
\nonumber
\eeq
This equation implies (\ref{qv}), due to the relation $\sum_n \delta v_n=0$
and equations (\ref{110}). The lemma is proved.

Representation of  symplectic forms as a residue at infinity of the
spectral curve of meromorphic differential
$d\Omega$ in the right hand side of (\ref{omega})
allows to find Darboux coordinates. These Darboux coordinates are connected
with residues of the differential at finite points.

\begin{lem} Let $R=\prod_s(E-\g_s)$ be the monic polynomial with zeros
at the poles of the Baker-Akhiezer function $\psi^{\s}$. Then
the following identity holds
\beq
{dw\over <\psi^+\psi>}={RdE},
\nonumber
\eeq
\end{lem}
{\it Proof.} The equation $(L-E)\psi=0$ implies that $(dL-dE)\psi=(E-L)d\psi$.
Therefore,
$$<\psi^+(dL-dE)\psi>=<\psi^+(E-L)d\psi>=0, $$
and we obtain the equality
$$
dE<\psi^+\psi>=<\psi^+dL\psi>=c_{N-1}dw\left(\psi^+_{N-1}\psi_0\right).
$$
From (\ref{ps}) we have that $\psi_{N-1}^+=c_{N-1}^{-1}R^{-1}$. 
Lemma is proved.

This result implies that the differential $d\Omega$, besides $E=\infty$, may have poles
at spectral points $E_k$, at poles $\g_s$ of $\psi^+$, and at zeros $z_i$ of $dw$.
(The last poles are due to (\ref{de}).)

First of all, let us show that $d\Omega$ has no poles at $E_k$.
At these points $L=L_0$, and therefore, is symmetric. Using gluing conditions
we obtain
$$
<\psi^+\delta L \wedge \delta \psi>+
<\delta\psi^+\wedge\delta L \psi>|_{w=0}=
<\psi^T\delta L_0 \wedge \delta \psi>+
<\delta\psi^T\wedge\delta L_0 \psi>=0,
$$
due to skew-symmetry of the wedge product.

At $E=\g_s$ the first term of $d\Omega$ does not contribute to residue because poles of
$\psi^+$ cancel with zero of $R$. Nontrivial residue is due to second order
pole of $\delta \psi^+$. We have
\begin{eqnarray}
\R_{\g_s}
{<\delta\psi^+\wedge\delta L \psi>\over <\psi^+\psi>} d\ln w&=&
\delta \ln w(\g_s)\wedge
\left[{<\psi^+\delta L \psi>\over <\psi^+\psi>}\right]_{E=\g_s}
\nonumber\\
{}&=&
\delta \ln w(\g_s)\wedge \delta E(\g_s).
\label{200}
\end{eqnarray}
In the last equality we use the well-known formula for variation of eigenvalue
$$ <\psi^+\delta L\psi>=\delta E<\psi^+\psi>,$$
which immediately follows from the equation
$$<\psi^+(\delta L-\delta E)\psi>=<\psi^+(L-E)\delta\psi>=0.$$
Consider now critical points $z_i$ of the polynomial $w(E): \ dw(z_i)=0.$
Formula (\ref{de0}) for $\delta \psi$ implies
\beq
\R_{z_i}
{<\psi^+\delta L \wedge\delta \psi>\over <\psi^+\psi>}\  d\ln w=
\R_{z_i}
{\left[<\psi^+\delta L d\psi>\wedge \delta E(w)\right]\ d\ln w
\over dE <\psi^+\psi>}
\nonumber
\eeq
Due to skew-symmetry of the wedge product, $\delta L$ in the last formula
may be replaced by $(\delta L-\delta E)$.
The equations
$\psi^+(\delta L-\delta E)=-\delta \psi^+(L-E)$ and
$(L-E)d\psi=-(dL-dE)\psi$ imply
$$
<\psi^+(\delta L-\delta E) d\psi>=
<\delta \psi^+(dL-dE) \psi>\ .
$$
Note that $dL(z_i)=0$. Hence,
$$
\R_{z_i}
{<\psi^+\delta L \wedge\delta \psi>\over <\psi^+\psi>}\  d\ln w=
-\R_{z_i}
{\left(<\delta \psi^+\psi>\wedge \delta E\right) \  d\ln w\over <\psi^+\psi>}\ .
$$
In the same way we obtain
$$
\R_{z_i}
{<\delta \psi^+\wedge \delta L \psi>\over <\psi^+\psi>}\  d\ln w=
-\R_{z_i}
{\left(\delta E\wedge <\psi^+\delta \psi>\right)\  d\ln w\over <\psi^+\psi>}\ .
$$
Taking a sum of these equalities we get
$$
\R_{z_i} d\Omega=
\R_{z_i}
{\left(<\psi^+\delta \psi-\delta \psi^+\psi>\wedge \delta E\right) \
d\ln w\over <\psi^+\psi>}\ .
$$
Besides poles at the points $z_i$, the differential at the right hand side
of the last equation has poles at the points $\g_s$, only. Indeed,
it has no pole at the infinity, because due to constraint $\sum_n\delta v_n=0$
we have $\delta E=O(k^{-1})$. There is no pole of the differential at
zeros $E_k$ of $w(E)$ due to gluing conditions.

Hence, we get
\beq\label{201}
\sum_i \R_{z_i}d\Omega=-\sum_{s}\R_{\g_s}
{\left(<\psi^+\delta \psi-\delta \psi^+\psi>\wedge \delta E\right) \
d\ln w\over <\psi^+\psi>}=\sum_s\delta\ln w(\g_s)\wedge \delta E(\g_s)
\eeq
Sum of all the residues of $d\Omega$ equals to zero. Therefore, using
(\ref{200}, \ref{201}) we obtain
\begin{lem}
Let $\omega$ be a two-form given by the formula (\ref{omega}). Then
\beq
\omega=2\sum_{s=1}^{N-1}\delta E(\g_s)\wedge \delta \ln w(\g_s).
\nonumber
\eeq
\end{lem}
Finally, combining results of Lemma 4 and 6 we obtain
\begin{th} Poles $\g_s=E(\g_s)$ of the Baker-Akhiezer function and
evaluation $\ln w(\g_s)$ are Darboux coordinates
for reduction of the canonical symplectic form on the leaf ${{\cal M}_I}$,
defined by the constraint $\sum_n v_n=I$:
$$\left(\sum_{n}\delta v_n\wedge \delta q_n\right)_{{\cal M}_I}=
\sum_{s=1}^{N-1}\delta \ln w(\g_s)\wedge \delta \g_s.
$$
\end{th}
From (\ref{P}) we have $w(\g_s)=\prod_k(\g_s-E_k)$. Therefore,
\beq\label{triv}
\sum_{s=1}^{N-1}\delta \ln w(\g_s)\wedge \delta \g_s=
\sum_{s=1}^{N-1}\sum_{k=1}^N{\delta E_k\wedge \delta \g_s\over E_k-\g_s}=
\sum_{k=1}^{N} \delta \ln r_k\wedge \delta E_k.
\eeq
where $r_k=R(E_k)=\prod_s(E_k-\g_s)$ are variables used in the previous
section. From the formulae
$\varphi_0(E_k)\psi_{N-1}(E_k)=1$ and $\varphi_0(E_k)=c_{N-1}r_k,\quad k=1,...,N;$
we obtain
$$
\sum_{k=1}^{N} \delta \ln r_k\wedge \delta E_k =
\sum_{k=1}^{N} \delta \ln \varphi_0(E_k) \wedge \delta E_k=
-\sum_{k=1}^{N} \delta \ln \psi_{N-1}(E_k) \wedge \delta E_k
$$
Note that in all  formulae the variables $E_k$ are subject to the constraint
$\sum_k E_k=\sum_n v_n=I$. The last representation for the symplectic structure in
terms of first/last components of normalized eigenfunctions and eigenvalues of $L_0$
is well-known, see \cite{MO}.

We would like to emphasize that the form in  (\ref{triv})
can be seen as a limit
$\Lambda\to 0$ of the  formula for  action-angle variables
for the periodic Toda lattice. To see this, let us choose
$E_1,\ldots E_{N-1}$, as
independent variables, then
\beq
\sum_{k=1}^{N} \delta \ln r_k\wedge \delta E_k=
\sum_{k=1}^{N-1} \delta \left(\ln r_k-\ln r_N\right)\wedge \delta E_k.
\eeq
If we choose first $(N-1)$ cuts of the hyperelliptic curve as a basis
of $a$-cycles, then $E_k$ are identified with limit of the action
variables for the periodic Toda lattice:
$$
E_k=\lim_{\Lambda\to 0}\ \ {1\over 2 \pi i} \oint_{a_k}Ed\ln w_0\ ,
$$
where $w_0=\Lambda w$ from  (\ref{sp0}). Limit of the normalized holomorphic
differential $d\Omega_k$ on the spectral curve of the periodic
Toda lattice is equal to
$$
\lim_{\Lambda\to 0}\ \ d\Omega_k={dE\over(E-E_k)}-{dE\over(E-E_N)}\ .
$$
Therefore, the variables $(\ln r_k-\ln r_N)$ are identified with
a limit of the angle variables $\phi_k$ in the periodic case, which
are result of the Abel transform of poles of the Floquet-Bloch
solution
$$
\ln r_k-\ln r_N=\lim_{\Lambda\to 0} \sum_{s}\int^{\g_s}d\Omega_k=
\lim_{\Lambda\to 0} \ \ \phi_k.
$$

\section{2D open Toda lattice}

Equations of motion of 2D open Toda lattice are equations for $N$
unknown functions $q_n(x,t)$. In light-cone coordinates $\xi=x+t, \ \eta=x-t$
they have the form
$$
\p^2_{\xi\eta}q_n=e^{q_{n-1}-q_n}-e^{q_n-q_{n+1}},\ n=1,\ldots, N-2,
$$
\beq
\p^2_{\xi\eta}q_0=-e^{q_0-q_{1}},\
\p^2_{\xi\eta}q_{N-1}=e^{q_{N-2}-q_{N-1}}\ .\label{2d}
\eeq
We consider the Goursat initial value problem with
initial data given on characteristics $\p_{\xi} q_n(\xi,0)$ and $q_n(0,\eta)$.    

In two-dimensional case an analog of the Lax representation is the
zero-curvature equation $[\p_{\xi}-U_0,\p_{\eta}-V_0]=0$, where
$U_0$ and $V_0$ are finite-dimensional matrices
\beq\label{UV}
U_0=\left[\begin{array}{ccccc}
v_0      & 1       & 0         &\cdots        & 0\\
0      & v_1       & 1       &\cdots        & 0\\
\cdot        &  \cdot    & \cdot     &\cdot         & \cdot\\
0    &  \cdots         & 0         & v_{N-2}     & 1\\
0       & \cdots    & 0         & 0      & v_{N-1}
\end{array}\right],\ \
V_0=\left[\begin{array}{ccccc}
0      & 0       & 0         &\cdots        & 0\\
c^2_0      &   0     & 0       &\cdots        & 0\\
\cdot        &  \cdot    & \cdot     &\cdot         & \cdot\\
0    &  \cdots         & c^2_{N-3}        & 0     & 0\\
0       & \cdots    & 0         & c^2_{N-2}      & 0
\end{array}\right]
\eeq
where, as before $c_n^2=e^{q_n-q_{n+1}}$ and $v_n=-\p_{\xi}q_n$.

Note, that for $N=2$ equations (\ref{2d}) imply for $\varphi=q_0-q_1$
the Liouville equation
$$
\p^2_{\xi\eta}\varphi=-2e^{\varphi}.
$$
Explicit solution for the Liouville equation which contains two arbitrary
was constructed by Liouville himself \cite{liouv}. 
There are many approaches that lead to generalization of this solution for 
the case of arbitrary $N$ (see, for example, \cite{lez,etin}).

The main goal of this section is to show that the general solution of the
open 2D Toda lattice can be obtained as a degenerate case of the construction
proposed in \cite{K4} for solving of periodic 2D Toda lattice. In \cite{K4}
it was shown that any local solution of periodic 2D Toda lattice
has an analog of the d'Alambert  representation as a composition of
functions depending on $\xi$  (or $\eta$), only. Nonlinear superposition
of such right- (or left-) moving waves is achieved with the help of
auxiliary {\it linear} Riemann-Hilbert problem. The corresponding matrix
Baker-Akhiezer function which solves the Riemann-Hilbert problem has
essential singularities at two points on complex plane of the spectral
parameter $w$. If we represent this plane as two-fold cover
with the help of the equation $w+\Lambda^2w^{-1}=E$
and take $\Lambda\to 0$, then the limit of the two-fold cover
becomes a singular curve, which can be seen as two copies of the plane
attached to each other at one point. In this limit, as we show the
Riemann-Hilbert problem reduces to a system of linear equations and
provides the general solution for 2D open Toda lattice.

Now let us provide details. Initial data which define the general solution
are arbitrary functions of one variable
$a_0(\xi),\ldots,a_{N-1}(\xi),\ b_0(\eta),\ldots,b_{N-2}(\eta)$
which can be identified with initial data for $q_n$ on the characteristics
$a_n(\xi)=v_n(\xi,0),\ \ b_n(\eta)=c_n^2(0,\eta)$.

Let $\Phi(\xi,w)$ and $\Phi^{\s}(\eta,w)$ matrix solutions
of ordinary differential equations
\beq\label{gem}
\left(\p_{\xi}-A(\xi,w)\right)\Phi(\xi,w)=0,\ \
\left(\p_{\eta}-B(\eta,w)\right)\Phi^{\s}(\eta,w)=0.
\eeq
normalized by the initial condition
\beq\label{n}
\Phi(0,w)=\Phi^{\s}(0,w)=1,
\eeq
Here $A$ and $B$ are the matrices
\beq\label{UVa}
A(\xi,w) =\left[\begin{array}{ccccc}
a_0      & 1       & 0         &\cdots        & 0\\
0      & a_1       & 1       &\cdots        & 0\\
\cdot        &  \cdot    & \cdot     &\cdot         & \cdot\\
0    &  \cdots         & 0         & a_{N-2}     & 1\\
w       & \cdots    & 0         & 0      & a_{N-1}
\end{array}\right],\ \
B(\eta,w) =\left[\begin{array}{ccccc}
0      & 0       & 0         &\cdots        & b_{N-1}w\\
b_0      &   0     & 0       &\cdots        & 0\\
\cdot        &  \cdot    & \cdot     &\cdot         & \cdot\\
0    &  \cdots         & b_{N-3}        & 0     & 0\\
0       & \cdots    & 0         & b_{N-2}      & 0
\end{array}\right]
\eeq
As before, we define $b_{N-1}$ by the formula
$b_{N-1}=\prod_{n=0}^{N-2}b_n^{-1}$.

Let $R(\xi,\eta)$ and $R^{\s}(\xi,\eta)$ be lower- and upper- triangular
matrices
$$R_{ii}=1,\ \ R_{ij}=0, i<j;\ \ R^{\s}_{ij}=0, \ i\geq j.$$
They are uniquely defined by the {\it gluing} condition at the point $w=0$
\beq\label{gl1}
\Psi(\xi,\eta,0)=\Psi^{\s}(\xi,\eta,0)
\eeq
for matrices
\beq\label{PF}
\Psi(\xi,\eta,w)=R(\xi,\eta)\Phi(\xi,w),\ \
\Psi^{\s}(\xi,\eta,w)=R^{\s}(\xi,\eta)\Phi^{\s}(\eta,w),\ \
\eeq
If diagonal matrix $H=H_{i}\delta_{ij}$, upper- and lower- diagonal matrices
$B_+, \ B_-$ with units on diagonals, are defined by the Borel decomposition
of the matrix
\beq\label{BO}
\Phi(\xi,0)\Phi^{\s}(\eta,0)^{-1}=B_-HB_+,
\eeq
then (\ref{gl1}) implies that $R=B_-^{-1},\  R^{\s}=HB_+$. In particular
we have $R^{\s}_{ii}=H_i$.

\begin{th}
The functions
$$q_n(\xi,\eta)=-\ln R_{nn}^{\s}(\xi,\eta)$$
solve the open 2D Toda lattice equations. They satisfy the initial conditions
$$\p_\xi q_n(\xi,0)=-a_n(\xi),\ \
q_{n}(0,\eta)-q_{n+1}(0,\eta)=\ln b_n(\eta).
$$
\end{th}
{\it Proof.}  Let us show first, that the matrix functions
$\Psi$ and $\Psi^{\s}$, satisfy linear equations
\beq\label{U}
(\p_\xi-U)\Psi=0, \ (\p_\xi-U_0)\Psi^{\s}=0,
\eeq
where $U_0=U_0(\xi,\eta,w)$ has the form (\ref{UV}) with $v_n=-\p_{\xi}q_n$
and $U=U_0+we^-$. Here $e^-$ is the matrix with the only non-vanishing entry
at the left lower corner: $e^-_{ij}=\delta_{i,N-1}\delta_{0,j}$.

The matrix $\Psi$ is non-degenerate. Therefore, its logarithmic derivative
$U=\p_{\xi}\Psi \Psi^{-1}$ is holomorphic function of $w$. Moreover, from 
(\ref{PF}) and condition that $R$ is a lower-triangle matrix it follows that
\beq\label{U11}
U=R_{\xi}R^{-1}+RAR^{-1}
\eeq
have the form $U=we^-+\tilde U$, where $\tilde U$ does not depend on $w$
and have vanishing entries over the first diagonal over the main diagonal
\beq\label{U13}
\tilde U_{i,i+1}=1,\ U_{ij}=0, \ j>i+1.
\eeq 
On the other hand, from (\ref{PF}) we have that 
\beq\label{U12}
U_0=\p_{\xi}\Psi^{\s}\left(\Psi^{\s}\right)^{-1}=R^{\s}\left(R^{\s}\right)^{-1}
\eeq 
is {\it upper-triangular} matrix with diagonal elements equal $-\p_{\xi}q_n$.
Gluing conditions (\ref{gl1}) imply
\beq
U(\xi,\eta,0)=\tilde U(\xi,\eta)=U_0(\xi,\eta)
\eeq
Therefore, combining the conditions (\ref{U13}) and the fact that $U_0$ is
upper-triangular we conclude that $U_0$ has the form (\ref{UV}).

In the similar way one proves that $\Psi$ and $\Psi^{\s}$, satisfy linear 
equations
\beq\label{V}
(\p_\eta-V_0)\Psi=0, \ (\p_\xi-V)\Psi^{\s}=0,
\eeq
where $V=V_0+w(e^-)^T$. Compatibility conditions of (\ref{U}) and (\ref{V})
imply that $q_n$ solves 2D Toda lattice equations.

At $\eta=0$ due to (\ref{n}) we have that matrices $R(\xi,0), R^{s}(\xi,0)$
are defined by the Borel decomposition of the matrix 
$$ \Phi(\xi,0)=R^{-1}(\xi,0)R^{\s}(\xi,0)$$
But this matrix is upper-triangular. Therefore, $R(\xi,0)=1$ and
$$U_0(\xi,0)=A(\xi).$$ In the same way we prove that $V_0(0,\eta)=B(\eta)$.
The last two equalities gives initial conditions on the characteristics for 
$q_n$. The theorem is proved.

We would like to emphasize that the matrix-functions 
$\Phi(\xi,0)$ and $\Phi^{\s}(\eta,0)$, which define through the Borel 
decomposition solutions of 2D Toda lattice, can be written explicitly.
From (\ref{UVa}) it follows that $\Phi^{\s}(\eta,0)$ is a 
lower-triangular matrix
\beq\label{fs1}
\Phi^{\s}_{ij}(\eta)=0, \ i<j,\ \Phi_{jj}^{\s}(\eta,0)=1,
\eeq
Using (\ref{fs1}) as initial condition for integration of equation (\ref{gem}),
$\p_{\eta}\Phi^{\s}_{kj}=b_k\Phi_{k-1,j}^{\s},$ we obtain that  
for $i>j$ entries of $\Phi^{\s}$ are given by multiple integrals
\beq\label{fs2}
\Phi^{\s}_{i,j}(\eta)=
\int_{0}^{\eta}\int_{0}^{\eta_{i-1}}\ldots
\int_{0}^{\eta_j}\prod_{k=1}^{i-j} b_{j+k}(\eta_{j+k-1})d\eta_{j+k-1}.
\eeq
In the same way we may right explicit formulae for the matrix $\Phi(\xi)$.
It is upper triangular and if we introduce functions
$$\tilde q_n(\xi)=\exp\left(-\int_{0}^{\xi}v_n(\xi')d\xi'\right),\ 
\tilde b_n=e^{\tilde q_n-\tilde q_{n+1}}$$
then
\beq\label{f1}
\Phi_{ij}(\xi)=0, \ i>j,\ \Phi_{ii}(\xi)=e^{-\tilde q_i(\xi)},
\eeq
and for $i<j$ its entries are given by multiple integrals
\beq\label{fs21}
\Phi_{i,j}(\xi)=
e^{-\tilde q_i(\xi)}
\int_{0}^{\xi}\int_{0}^{\xi_{i+1}}\ldots
\int_{0}^{\xi_j}\prod_{k=1}^{j-i} \tilde b_{j-k}(\eta_{j-k+1})d\eta_{j-k+1}.
\eeq
Let $M^{(n)}(\xi,\eta)$ be $(n+1)\times (n+1)$ matrices which are
upper left blocks 
\beq\label{M100}
M^{(n)}_{ij}=M_{ij}, \ 0\leq i,j\leq n,
\eeq
of the matrix 
\beq\label{M101} 
M(\xi,\eta)=\Phi(\xi,0)\Phi^{\s}(\eta,0)^{-1}.
\eeq
Then well-known formula for diagonal matrix of the Borel decomposition 
(\ref{BO}) 
$$
H_n= {D_n\over D_{n-1}},\ \ D_n=\det M^{(n)}, \ D_{-1}=1$$
together with Theorem 5 provide  an  explicit formula  for the solution
$$q_n(\xi,\eta) =-\ln {D_n\over D_{n-1}}$$
normalized by the condition
$$q_0(0,\eta)=0$$
Note that equations of 2D Toda lattice are invariant with respect
to the transformation $q_n\to q_n+f(\eta)$, where $f(\eta)$ is an arbitrary 
function. This transformation does not change initial data on characteristics.

\end{document}